\documentclass[lettersize,journal]{IEEEtran}
\usepackage{amsmath,amsfonts}
\usepackage{algorithmic}
\usepackage{algorithm}
\usepackage{array}
\usepackage[caption=false,font=normalsize,labelfont=sf,textfont=sf]{subfig}
\usepackage{textcomp}
\usepackage{stfloats}
\usepackage{url}
\usepackage{verbatim}
\usepackage{graphicx}
\usepackage{cite}
\usepackage{multirow}
\begin{document}

\title{Satellite-MEC Integration for 6G Internet of Things: Minimal Structures, Advances, and Prospects}

\author{Yueshan Lin, Wei Feng, \IEEEmembership{Senior Member, IEEE}, Yanmin Wang, Yunfei Chen, \IEEEmembership{Senior Member, IEEE}, Yongxu Zhu, \IEEEmembership{Senior Member, IEEE}, Ximu Zhang, Ning Ge, \IEEEmembership{Member, IEEE}
	
	\thanks{Yueshan Lin, Wei Feng, Ximu Zhang and Ning Ge are with the Department of Electronic Engineering, Tsinghua University, Beijing 100084, China (email: lin-ys17@tsinghua.org.cn, fengwei@tsinghua.edu.cn, zhangximu@tsinghua.edu.cn, gening@tsinghua.edu.cn).}
	\thanks{Yanmin Wang is with the School of Information Engineering, Minzu University of China, Beijing 100041, China (email: yanmin-226@163.com).}
	\thanks{Yunfei Chen is with the Department of Engineering, University of Durham, Durham DH1 3LE, U.K. (e-mail: Yunfei.Chen@durham.ac.uk).}
	\thanks{Yongxu Zhu is with the School of Engineering, University of Warwick, Coventry CV4 7AL, U.K. (email: yongxu.zhu@warwick.ac.uk).}
}

\maketitle

\begin{abstract}
The sixth-generation (6G) network is envisioned to shift its focus from the service requirements of human beings' to those of Internet-of-Things (IoT) devices'. Satellite communications are indispensable in 6G to support IoT devices operating in rural or disastrous areas. However, satellite networks face the inherent challenges of low data rate and large latency, which may not support computation-intensive and delay-sensitive IoT applications. Mobile Edge Computing (MEC) is a burgeoning paradigm by extending cloud computing capabilities to the network edge. By utilizing MEC technologies, the resource-limited IoT devices can access abundant computation resources with low latency, which enables the highly demanding applications while meeting strict delay requirements. Therefore, an integration of satellite communications and MEC technologies is necessary to better enable 6G IoT. In this survey, we provide a holistic overview of satellite-MEC integration. We first discuss the main challenges of the integrated satellite-MEC network and propose three minimal integrating structures. For each minimal structure, we summarize the current advances in terms of their research topics, after which we discuss the lessons learned and future directions of the minimal structure. Finally, we outline potential research issues to envision a more intelligent, more secure, and greener integrated satellite-MEC network.
\end{abstract}


\section{Introduction}
The past few years have witnessed the proliferation of intelligent Internet-of-Things (IoT) devices, such as wireless sensors, industrial robots and intelligent vehicles. With connection to the Internet, these IoT devices can enable a myriad of emerging applications (e.g., autonomous driving). It is estimated that the number of connected IoT devices will reach 30 billion by the end of 2025 \cite{intro01}. As a consequence, future sixth-generation (6G) networks will focus mainly on serving these intelligent IoT devices instead of human beings. Providing IoT devices with satisfactory services raises challenges for wireless system design. One major challenge is that a considerable part of the IoT devices are deployed in remote areas, such as oceans, deserts and forests, for environmental monitoring and resource exploitation. The harsh geographical conditions in these areas make it difficult or expensive to construct the traditional terrestrial infrastructures in fifth-generation (5G) networks. Besides, some IoT devices are required in disastrous areas, where the terrestrial infrastructures may suffer from serious damage. To address this, a non-terrestrial network via satellites and unmanned aerial vehicles (UAVs) may be used to complement the terrestrial network and fill the coverage gap.

Satellite communications are considered a promising solution to providing ubiquitous broadband Internet access at low cost \cite{intro02}. Geostationary earth orbit (GEO) satellite networks, which are traditionally used for satellite communications, have experienced rapid development in terms of providing high speed services for global users \cite{intro03}. Moreover, low earth orbit (LEO) constellation networks have attracted great attention due to their lower propagation latency and path loss \cite{intro04}. Several commercial projects of LEO satellite communication, such as OneWeb, Telesat, and Starlink, have been launched. Despite the many advantages, satellite networks also face their inherent challenges. Compared with terrestrial networks, satellite networks typically have lower data rates and larger latency. In computation-intensive applications, IoT devices may need to offload their data for cloud computing, due to their limited computing resources. Offloading through satellites can lead to a large delay, which is unacceptable for IoT devices that require delay-sensitive services.

To address these challenges, one promising direction is to enable edge intelligence to replace traditional cloud computing \cite{intro05}. The basic idea is to extend cloud computing capabilities to the network edge to enable artificial intelligence (AI) applications, so that the IoT devices will be endowed with low-latency data processing and decision-making capabilities. Mobile edge computing (MEC) technologies play an important role in the edge intelligence paradigm. In current 5G networks, MEC technologies have been used to enhanced the service quality for human beings. We envision that integrating satellite networks and MEC can better support IoT applications in remote or disastrous areas too \cite{intro06}.

Preliminary attempts have been made to integrate satellite communications and MEC technologies. For instance, Hewlett Packard Enterprise (HPE) partnered with National Aeronautics and Space Administration (NASA) to first launch computers to the International Space Station, namely the HPE SpaceBorne Computer, which managed to operate during its full time aboard. In addition, the cloud service providers ({\it e.g.}, Amazon, Microsoft and Google) have explored cloud-based ground stations which directly connect satellites with ground data centers. Despite these efforts, the design of an integrated satellite-MEC network is still an open issue.

Many studies have explored different aspects of the integrated satellite-MEC network. However, there is not yet a comprehensive review of the existing studies on satellite-MEC integration. In this paper, we introduce three minimal structures of the integrated satellite-MEC network. For each minimal integrating structure, we examine the existing studies based on their research topics, and discuss possible future research directions. We also provide a review of the studies that consider a combination of the minimal structures.

The rest of this paper is organized as follows. In Section II, we present the main challenges of satellite-MEC integration and summarize existing works that point out research directions, after which we introduce the three minimal structures of the integrated satellite-MEC network. In Sections III, IV and V, we discuss the three minimal structures. We summarize the advances on each structure based on their research topics, and further discuss the learned lessons and possible future directions. In Section VI, we review the studies that investigate a combination of the minimal structures. Finally, Section VII outlines open issues, and Section VIII draws the conclusion.

\IEEEpubidadjcol

\section{Minimal Integrating Structures}
To provide satisfactory services for IoT devices and enable new applications, an integrated satellite-MEC network is necessary. However, the integration of satellite networks and MEC has several main challenges as listed below \cite{mag00}.

\begin{itemize}
	\item {\bf Unique Characteristics of Service Requirements.} Compared with terrestrial networks, the integrated satellite-MEC network focuses more on IoT devices working in rural or disastrous areas. Their service requirements have several unique characteristics. First, the spatial distribution of IoT devices' service requirements is much sparser than the terrestrial network. Besides, the spatial distribution of these service requirements can vary significantly over time, in terms of service number, service type, {\it etc}. Moreover, the spatial and temporal distributions of these service requirements can be highly heterogeneous. These unique characteristics bring challenges to the integration.
	
	\item {\bf Limited Communication and Computing Capability.} The communication and computation resources of the integrated satellite-MEC network are quite limited. The communication resources are limited, because on the one hand, the limited bandwidth resources and long propagation distance result in low data rate and large latency. On the other hand, the limited orbit resources restrict the number of operating satellites. The computing resources are also limited, because MEC servers deployed on satellites or UAVs are restricted in terms of size, weight and energy, and radiation-hardened servers on satellites incur extra costs. The insufficient communication and computing resources make it difficult for the integrated satellite-MEC network to provide satisfactory services.
	
	\item {\bf Complexity of Matching Network Resources with Service Requirements.} Due to the limited communication and computing resources, a key issue in the integrated satellite-MEC network is to configure proper network resources to match the service requirements, in order to achieve higher resource utilization efficiency. This can be difficult because there exist hierarchical communication and computing resources in the integrated network, such as the communication links of different features and the multiple layers of edge servers. Moreover, the resources change dynamically due to the mobility of UAVs and LEO satellites, making the problem more complicated.
\end{itemize}

To tackle these challenges, some existing works have made great contributions in outlining research directions of the integrated satellite-MEC network. To improve users’ Quality-of-Service (QoS), Zhang \textit{et al.} \cite{mag01} investigated possible ways to implement MEC techniques in satellite-terrestrial networks. A dynamic network function virtualization technique was further proposed to unify the available network resources. Xie \textit{et al.} \cite{mag02} presented an architecture design of the satellite-terrestrial edge computing network based on certain design principles. After introducing the functional components of this architecture, the technical challenges of the network were discussed. Cassara \textit{et al.} \cite{mag03} envisioned integrating edge computing with LEO satellites, where the edge-cloud continuum concept was introduced. The authors further provided a reference scenario of the network architecture, and discussed the offloading strategies therein. Shang \textit{et al.} \cite{mag04} illustrated the computing architecture of satellite-aerial-ground integrated networks and presented cooperative computing schemes for various computing scenarios.

Our previous work \cite{mag00} discussed typical use cases and main design challenges of the integrated satellite-MEC network. We proposed three minimal integrating structures of the integrated network and discussed their characteristics and problems, and further designed an on-demand network orchestration framework. Specifically, the three minimal structures were the Computing-in-Forward-link (CIF) structure, the Computing-on-Orbit (COO) structure, and the Computing-after-Feeder-link (CAF) structure. The complicated integrated satellite-MEC network could be viewed as a nonlinear orchestration of these three minimal structures, which provided a new and simple perspective on the integrated network's system design. In the following, we will categorize relevant studies according to the three minimal structures, and conduct a comprehensive review of their research problems.

\begin{figure*}[t]
	\centering
	{\includegraphics[height=2.2in]{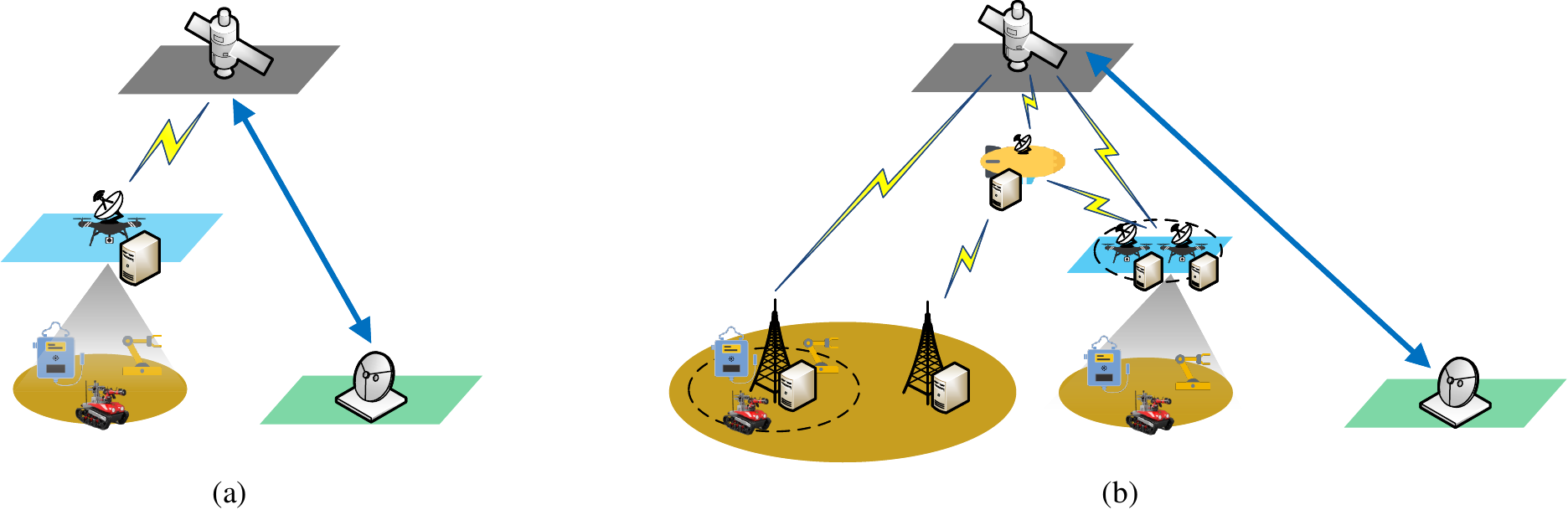}%
		\label{CIF basic}}
	\caption{(a) Illustration of the basic CIF structure \cite{mag00}. (b) Illustration of one possible extension of the CIF structure.}
\end{figure*}

\section{Computing-in-Forward-link Structure}
The first minimal structure is the CIF structure. As shown in Fig. 1(a), this minimal integrating structure consists of an aerial platform (AP) equipped with an MEC server, a satellite, a gateway and multiple IoT devices. The AP can be a UAV or a high-altitude platform (HAP). This minimal structure can be extended by considering multiple APs, and by including ground base stations with MEC equipment in the network, as shown in Fig. 1(b).

By deploying MEC servers in proximity to IoT devices, the CIF structure is suitable for local-area applications with ultra-low latency requirements. However, the central processing unit (CPU) capability of an on-board MEC server is restricted since APs are limited in size and energy, which should be considered in the system design.

Researches on the CIF structure mainly focus on two topics: computation offloading and content delivery.


\subsection{Computation Offloading}
Computation task offloading is a basic service provided by integrated satellite-MEC networks. For the CIF structure, an important problem is to properly allocate different users’ computation tasks to MEC servers. To solve the problem, it is necessary to comprehensively consider the characteristics of computation tasks (e.g., delay requirement, input data size), as well as the heterogeneous communication and computation resources in the network.

In \cite{J-077}, the CIF network consisted of a satellite and a multi-antenna access point with MEC capabilities. The access point worked in full-duplex mode, and thus the computation results could be transmitted back to users in real time. To improve the offloading data rate, the authors investigated users’ task offloading decision and resource allocation jointly. The authors of \cite{J-016} considered utilizing satellite and multiple UAVs in the network, where the UAVs were equipped with MEC servers to provide computation services. In this paper, the task offloading decision was jointly designed with the allocation of user power, bandwidth and computing resources, and the aim was to minimize the total energy cost in the system. The authors of \cite{J-016} proposed an algorithm based on double deep Q-learning as a solution. In \cite{C-131}, multiple UAVs and multiple ground base stations, all equipped with MEC servers, were used to provide edge computing services. An LEO satellite was connected to both the UAVs and the base stations for backhaul transmission. A joint UAV placement and task offloading decision problem was considered in the network to maximize the overall profit of the MEC service provider, which was determined by the number of finished computation tasks and the energy consumption of MEC servers. The authors provided a two-stage algorithm to solve this problem.

\subsection{Content Delivery}
In addition to computation offloading, another important service considered in the CIF structure is the delivery of bandwidth-demanding application data, such as high-resolution video streaming. Specifically, the broadcasting/multicasting capability of satellite communication enables content delivery to multiple network locations, where the data can be prestored in MEC servers in proximity to users. However, the system design still faces multiple challenges due to the limited network resources and long latency, and thus multiple studies have been conducted toward efficient content delivery.

\begin{table*}[t]
	\caption{Summary of Advances on the CIF structure}
	\centering
	\renewcommand{\arraystretch}{2}
	\begin{tabular}{|p{0.65in}|p{0.3in}|p{1.2in}|p{1in}|p{3in}|}
		\hline
		\textbf{Theme} & \textbf{Ref.} & \textbf{Network architecture} & \textbf{Design objective}  & \textbf{Proposed solution} \ \\ \hline
		\multirow{3}{0.5in}{Computation offloading} & \cite{J-077} & A satellite and a ground base station & Task offloading rate & Design users' task offloading decision and resource allocation, and provide a solution by decomposing the problem into two sub-problems. \\ \cline{2-5}
		~ & \cite{J-016} & A satellite and multiple UAVs & Energy & Jointly optimize the users' task offloading decision and resource allocation and propose a scheme based on double deep Q-learning. \\ \cline{2-5}
		~ & \cite{C-131} & A satellite, multiple UAVs and multiple ground base stations & Profit of MEC service provider & Jointly optimize the users' task offloading decision and UAV placement and propose a two-stage algorithm. \\ \hline
		\multirow{3}{0.5in}{Content delivery} & \cite{C-228} & MEC-enabled radio access network (RAN) with satellite backhaul & / & Investigate two use cases for popular and personalized content delivery. \\ \cline{2-5}
		~ & \cite{C-197} & MEC-enabled RAN with satellite and terrestrial backhaul & / & Propose a content delivery strategy to achieve optimal traffic distribution among satellite and terrestrial backhaul links. \\ \cline{2-5}
		~ & \cite{C-123} & MEC-enabled RAN with satellite backhaul & / & Propose a SR-based adaptive video streaming scheme.  \\ \hline
	\end{tabular}
	\label{CIF}
\end{table*}

The authors of \cite{C-228} proposed a network architecture where a CIF satellite-MEC network was utilized to support mobile video delivery. In this network, the authors investigated two use cases to enhance the users’ Quality of Experience (QoE). One use case used utilizing satellite communications to pre-populate video content to MEC servers at different locations based on the predictive content popularity. The other use case prefetched video content segments to the MEC servers, in order to overcome the long propagation latency of satellite links. In \cite{C-197}, a similar CIF structure was considered, except that both terrestrial and satellite backhaul links were included. The MEC server selected a backhaul link for each enhancement layer of the video, based on the playout buffer size. The authors proposed a content delivery strategy to achieve optimal traffic distribution among the backhaul links. The authors of \cite{C-123} proposed a super-resolution-based (SR-based) adaptive video streaming scheme in a CIF satellite-MEC network. Specifically, this SR-based method transmitted low-resolution images through the satellite links to overcome the limited transmission rate. The MEC server provided the computation resources necessary to run a deep neural network to reconstruct low-resolution images to high-resolution images. Table \ref{CIF} gives a summary of all these works.

\subsection{Lessons Learned and Future Directions}
Existing works on CIF structure mainly focus on two aspects. The first is the task offloading decision problem in computation offloading, and the second is content delivery. Possible new research directions are listed as follows.

First, new design objects can be considered for better system performance. For instance, the studies focused on task offloading decision considered energy cost or other indicators as design objectives, while the CIF structure mainly focuses on delay-sensitive services. Future studies can consider task latency as the design objective.

Besides, there are also new topics to be discussed. For instance, an important problem is to decide the on-board MEC capability of HAPs or UAVs. For this problem, the number of service requirements and the energy consumption are two most important factors to be considered. Since the duration of an HAP’s flight can be months, configuring the MEC capability on HAPs is a large-time scale problem. In this case, the design of the MEC capability may be based on the average service requirement number, which can be estimated by analyzing historical data. For the UAV case, however, the duration of one flight is only a few hours. In this case, the MEC capability configured for UAVs can be further optimized based on more specific information. For instance, the large-scale channel state information (CSI) during the UAV’s flight can be conveniently acquired from a pre-established database, referred to as a radio map \cite{radio map}. This external information can assist the on-board MEC capability configuration. Therefore, a new framework for medium-timescale network adjustment based on external information needs to be introduced, which can be further investigated.

In addition, the task pre-processing problem can be considered in this basic structure. Specifically, for computation tasks that have a huge input data size, the MEC servers can pre-process the task to compress the data size before offloading it to the cloud. This problem can be further discussed.

\begin{figure*}[t]
	\centering
	{\includegraphics[height=2.5in]{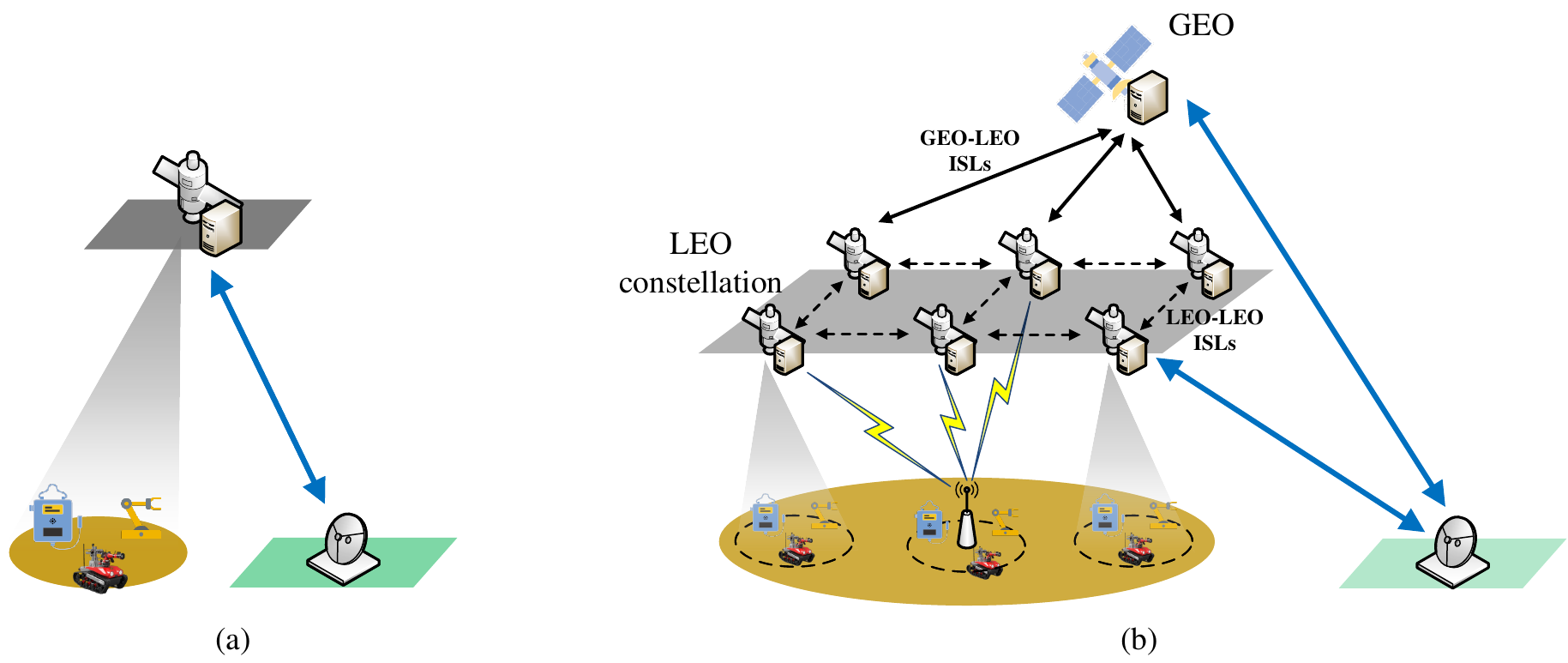}%
		\label{COO basic}}
	\caption{(a) Illustration of the basic COO structure \cite{mag00}. (b) Illustration of one possible extension of the COO structure.}
\end{figure*}

\section{Computing-on-Orbit Structure}
The second minimal structure is the COO structure. This minimal integrating structure is composed of a satellite equipped with an MEC server, a gateway and multiple IoT devices, as shown in Fig. 2(a). There are several variants of this minimal structure. For instance, the space segment can be a constellation of LEO satellites. In addition, a GEO satellite and an LEO constellation can coordinately provide edge computing services, as shown in Fig. 2(b).

Since satellites provide much larger coverage than APs, the COO structure can well support wide-area applications with low latency requirements. However, the minimal structure's system design faces multiple challenges due to the harsh space environment and the limitations of satellites in terms of size, weight and energy.

Researches on the COO structure mainly focus on three topics, namely MEC server placement, service placement and computation offloading.

\subsection{MEC Server Placement}
Since there can be multiple satellites on different orbits in space, the first problem of the COO structure is to determine on which satellites the MEC servers are placed. The harsh radiation environment in space can cause single-event effects, leading to abrupt changes in circuits and eventually inducing data losses or errors \cite{radiation-1}. Besides, the low temperature and vacuum in space could also cause damage to electronic devices \cite{radiation-2}. Therefore, placing MEC servers on satellites requires hardening measures for servers, which could incur extra costs. On the other hand, the satellites without MEC server equipment may need better inter-satellite links (ISLs) to offload their tasks. This leads to a tradeoff that requires careful consideration. Moreover, the temporal and spatial distributions of service requirements should also be considered, which makes the problem more complicated.

\begin{table*}[t]
	\caption{Summary of Advances on the COO Structure}
	\centering
	\renewcommand{\arraystretch}{2}
	\begin{tabular}{|p{0.65in}|p{0.3in}|p{1.3in}|p{1in}|p{2.9in}|}
		\hline
		\textbf{Theme} & \textbf{Ref.} & \textbf{Network architecture} & \textbf{Design objective}  & \textbf{Proposed solution} \ \\ \hline
		\multirow{2}{0.5in}{MEC server placement} &  \cite{C-060} & LEO constellation with ISLs & Number of placed servers & Design the MEC server placement method by modeling the LEO constellation as a 2D torus network and proposing an algorithm based on the $d$-hops placement method. \\ \cline{2-5}
		~ & \cite{C-096} & LEO constellation with ISLs & Latency & Design the MEC server placement and the association of MEC-enabled satellites and other satellites by a heuristic scheme based on the genetic algorithm. \ \\ \hline
		\multirow{2}{0.5in}{Service placement} & \cite{J-043} & LEO constellation & Service coverage and robustness & Optimize the service placement by proposing a Lyapunov optimization-based online service placement scheme. \\ \cline{2-5}
		~ & \cite{C-115} & LEO constellation with ISLs & Service satisfactory rate and ISL costs & Jointly optimize the service placement and service request scheduling scheme by modeling it as a mixed-integer linear programming problem to provide a solution. \\ \hline
		\multirow{10}{0.5in}{Computation offloading} & \cite{J-064} & LEO constellation & Energy & Design the task offloading scheme by proposing a distributed algorithm based on the alternating direction method of multipliers. \\ \cline{2-5}
		~ & \cite{C-054} & LEO constellation & Latency and energy & Jointly optimize the task offloading decision and the bandwidth and computation resource allocation, and solve the problem by decomposing it into two sub-problems.  \\ \cline{2-5}
		~ & \cite{J-101} & LEO constellation & Latency and energy & Optimize the task offloading, propose a game-theoretic approach to solve the problem and prove that the Nash equilibrium exists. \\ \cline{2-5}
		~ & \cite{J-058} & LEO constellation & Energy & Jointly optimize the task offloading decision and the power, bandwidth and computation resource allocation, and solve the problem by decomposing it into two sub-problems.  \\ \cline{2-5}
		~ & \cite{J-Sup-1} & LEO constellation with ISLs & Latency and energy & Develop a novel task allocation algorithm based on the greedy strategy. \\ \cline{2-5}
		~ & \cite{C-141} & LEO constellation and GEO with ISLs & Latency & Propose a task scheduling algorithm based on dynamic priority queue. \\ \cline{2-5}
		~ & \cite{C-042} & LEO constellation and GEO with ISLs & Latency and energy & Jointly optimize the task offloading decision and communication resource allocation, and propose an improved two-sided many-to-one matching game algorithm to solve the problem. \\ \cline{2-5}
		~ & \cite{C-089} & An LEO satellite and a ground base station & Profit of MEC service provider & Jointly optimize the task offloading decision and the communication and computation resource allocation, and solve the problem by decomposing it into two sub-problems. \\ \cline{2-5}
		~ & \cite{C-100} & LEO constellation and a UAV & Latency & Design the task offloading decision by proposing a curriculum learning-multi-agent deep deterministic policy gradient approach. \\ \cline{2-5}
		~ & \cite{C-019} & An LEO satellite and multiple UAVs & Latency and energy & Jointly design the task offloading decision and UAV trajectory by proposing a multi-agent reinforcement learning based algorithm. \\ \hline
	\end{tabular}
	\label{COO}
\end{table*}

In this context, both \cite{C-060} and \cite{C-096} explored the server placement problem in a COO network with LEO constellation in space. By modeling the LEO constellation as a two-dimensional torus network, the authors of \cite{C-060} aimed to place a minimum number of servers so that every satellite can access a server within a threshold distance. To achieve optimal server placement, an algorithm based on the $d$-hops placement method was proposed. On the other hand, the authors of \cite{C-096} focused on computation latency and considered two server placement problems. The first problem aimed to minimize the task response delay at a given snapshot, while the second one aimed for the average response delay for an entire time period. A heuristic scheme based on the genetic algorithm was proposed to solve both problems. The proposed scheme yielded a performance gain over traditional schemes since it took the temporal and spatial characteristics of LEO satellite networks into consideration.


\subsection{Service Placement}
The execution of a computation task requires not only computation resources but also a set of codes and related libraries/databases. The MEC server can pre-store the code and databases of certain services, which is referred to as service placement. Therefore, the next problem of the COO structure is the service placement decision of the satellite-based MEC servers. For this problem, it is of great importance to consider how different types of service requirements are distributed spatially.

In \cite{J-043}, the authors considered a COO system, where a constellation of satellites each equipped with an MEC server provided computing services. The service placement problem was investigated to maximize the robustness aware service coverage of the system. Specifically, the problem aimed to increase the user request number that can access the service, as well as the user request number that can access more than one service copy deployed on different satellite-based servers. The authors proposed an online service placement algorithm based on Lyapunov optimization and Gibbs sampling to give a near-optimal solution. The authors of \cite{C-115} further extended the system in \cite{J-043} by considering ISLs among LEO satellites. The joint service placement and service request scheduling problem was investigated, which aimed to reduce unsatisfactory service requests while minimizing the ISL transmission cost. The authors modeled it as a mixed-integer linear programming problem and provided a solution, which showed a better performance than the greedy methods.

\subsection{Computation Offloading}
For the COO structure, in addition to MEC server placement and service placement, another important problem is to properly offload users’ computation tasks to MEC servers. To solve this problem, it is necessary to comprehensively consider the characteristics of computation tasks (e.g., delay requirement, input data size), as well as the heterogeneous communication and computation resources in the network.

To implement computation offloading in the COO network, many existing works considered utilizing an LEO constellation in space, where each LEO satellite was equipped with an MEC server. In \cite{J-064}, \cite{C-054} and \cite{J-0417-006}, each user was associated with at most one satellite. In \cite{J-064}, the task offloading decision problem was considered to achieve minimum energy consumption of local and edge computing. A distributed algorithm based on the alternating direction method of multipliers was proposed, which approximated the optimal solution with low computational complexity. Adopting the same user association method, \cite{C-054} jointly optimized the task offloading decision and the bandwidth and computation resource allocation. To minimize the weighted sum of the energy consumption and task delay costs, the authors proposed an algorithm based on problem decomposition. Since service placement is preliminary to the computation offloading process, the authors of \cite{J-0417-006} considered the task offloading problem jointly with the service placement problem. For the minimization of task execution delays, the authors jointly optimized the service placement, task offloading decision and resource allocation of the system. An LDD-based algorithm was proposed to obtain the closed-form optimal solution, and a heuristic algorithm was also proposed to find an efficient solution with low complexity. In \cite{J-101} and \cite{J-058}, the users could offload their computation tasks to multiple satellites simultaneously. The authors of \cite{J-101} optimized the offloading decision to minimize the weighed sum of the average task response time and the average task energy consumption. The authors proposed a game-theoretic approach to solve this problem, which reached the Nash equilibrium in an iterative manner. In \cite{J-058}, joint optimization of task offloading decision and resource allocation was considered in the system. The aim was to minimize the total energy consumption of local and edge computing. The authors proposed a novel algorithm which decomposes the problem into two sub-problems and solves them respectively. \cite{J-Sup-1} further included ISLs in their considered system model. Specifically, users’ computation tasks were first offloaded to an access satellite, and then they could be forwarded to other satellites through ISLs for execution. The authors proposed a novel task allocation algorithm based on the greedy strategy to optimize the task offloading decision. The algorithm also focused on average delay and energy consumption reduction, and it showed a performance gain over the double edge computation offloading algorithm. In \cite{C-0417-023}, a special system model was considered where the computation task data were generated from source satellites ({\it e.g.}, Earth observation satellites) and offloaded to satellites with MEC equipment for edge computing. For energy consumption minimization, the task offloading decision and the communication and computation resource allocation were jointly optimized. The authors divided the original optimization problem into two sub-problems and applied successive convex approximation method to design an iterative algorithm.

Additionally, some studies considered a more complicated double-layer architecture of LEO and GEO satellites in space. In \cite{C-141}, each LEO satellite was equipped with an MEC server and executed the offloaded tasks, while the GEO satellites managed and coordinated the satellite MEC resources. To achieve task delay minimization, a scheduling algorithm based on dynamic priority queue was proposed to solve the task offloading decision problem. In \cite{C-042}, the computation tasks could be executed at LEO satellites or GEO satellites. With both latency and energy costs considered, the authors jointly optimized the task offloading decision and communication resource allocation. An improved two-sided many-to-one matching game algorithm was proposed to solve the problem.

Moreover, the combination of the COO structure with ground or aerial networks was investigated. The authors of \cite{C-089} considered a combined terrestrial-MEC and satellite-MEC network, where an LEO satellite provided edge services in space. In the system, the task offloading strategy and the resource allocation of the satellite were jointly considered, aimed at maximizing the profit of the MEC service provider. The proposed algorithm decomposed the problem into two sub-problems and produced a solution. Focused also on terrestrial-MEC and satellite-MEC combinations, the authors of \cite{J-0417-009} further considered a system model with multiple base stations. To minimize the total energy consumption, the task offloading decision was jointly optimized with the computing resource allocation. The authors adopted the classic alternating optimization method for decomposing the original problem and then solved each sub-problem using low-complexity iterative algorithms. The authors of \cite{C-100} considered combining aerial-MEC and satellite-MEC, where users could offload computation tasks to the LEO satellite or to a UAV flying on a predetermined trajectory. The optimization of task offloading decision was conducted to lower the time-averaged task execution latency. To learn the near-optimal offloading strategy, a curriculum learning-multi-agent deep deterministic policy gradient approach was proposed. In \cite{C-019}, the scenario of multiple UAVs and an LEO satellite was further considered, each equipped with an MEC server. The authors jointly optimized the task offloading decision and the UAV trajectory to for latency and energy cost minimization. A multi-agent reinforcement learning based task offloading algorithm was proposed to solve the problem. The authors of \cite{J-0417-013} further considered a system model with multiple LEO satellites and UAVs, equipped with MEC servers, to process or cache users' tasks. The task offloading decision problem was investigated to minimize the energy consumption for task execution. The authors employed a constrained Markov decision process to formulate the task offloading decision problem and further devised a deep reinforcement learning-based algorithm to solve the problem. Table \ref{COO} summarizes and compares these works.

\subsection{Lessons Learned and Future Directions}
There are three main research topics: deciding on which satellites the MEC servers are deployed, deciding the service placement of each server, and in a specific computation offloading application, deciding which server each task is offloaded to. Most works have focused on the last topic to investigate a joint task offloading decision and resource allocation problem. Many research gaps remain.

First, more realistic scenarios and environments should be considered. For instance, the severe electromagnetic radiation in space can influence satellite-based server performance and even cause damage. Therefore, the servers need to be radiation-hardened, which impacts the computation performance. Few existing works have considered this factor. Besides, the energy supply of satellites relies heavily on solar power, which can be inconsistent. This also influences satellite-based servers' performance. In future works, these factors should be taken into consideration to obtain more persuasive results.

In addition, a more complicated system model can be investigated. For instance, for the MEC server placement problem, MEC servers can also be placed on GEO satellites in addition to LEO satellites. With their inherently large coverage, the MEC-enabled GEO satellites can not only provide edge computing services for ground users, but also orchestrate the communication and computing resources for LEO satellites which enables better coordination in the system. This idea has been mentioned in \cite{C-141} and can be further investigated. 

Another research direction is to choose proper design objectives. We take the service placement problem as an example. In existing studies, the design objective of \cite{C-115} is service coverage and service robustness, while \cite{J-043} aimed to minimize the service satisfactory rate and ISL cost. In the future, novel design objectives should be considered to better describe the performance of service placement.

Moreover, some new research topics can be explored. One important instance is the MEC server activation problem. Due to the limited energy on satellites, adopting a full-on mode for MEC servers may be impractical. Therefore, it is important to decide which of the servers should be activated, in order to satisfy the service requirements and save the energy costs. Different from MEC server placement which is adjusted at similar timescales as infrastructure changes ({\it e.g.}, months), MEC server activation is often adjusted every few hours or minutes. Network adjustments at this timescale have yet to be investigated. Therefore, a novel network architecture that enables on-demand network adjustment at such a medium timescale needs to be considered \cite{mag00}.

For computation offloading, there are also new research topics to be considered. For instance, the scenario of multiple MEC servers executing a single complicated task can be considered. To make this happen, multiple satellites need to provide edge computing services in a coordinated manner. The authors of \cite{C-102} proposed an on-orbit federated learning system, where LEO satellites serve as local servers and a medium earth orbit (MEO) satellite serves as the global server. Further research can be conducted on this topic. Besides, the handover problem of satellite-based MEC servers can be considered. After the edge server finishes computation, the results need to be transmitted back to the user. This can be difficult due to the mobility of LEO satellites. Many existing works tackle this problem by setting a computation time constraint. However, this might not work when the offloading task is computationally intensive. In that case, the handover of computation results through ISLs is necessary, which can be further investigated.

\begin{figure*}[t]
	\centering
	{\includegraphics[height=2.2in]{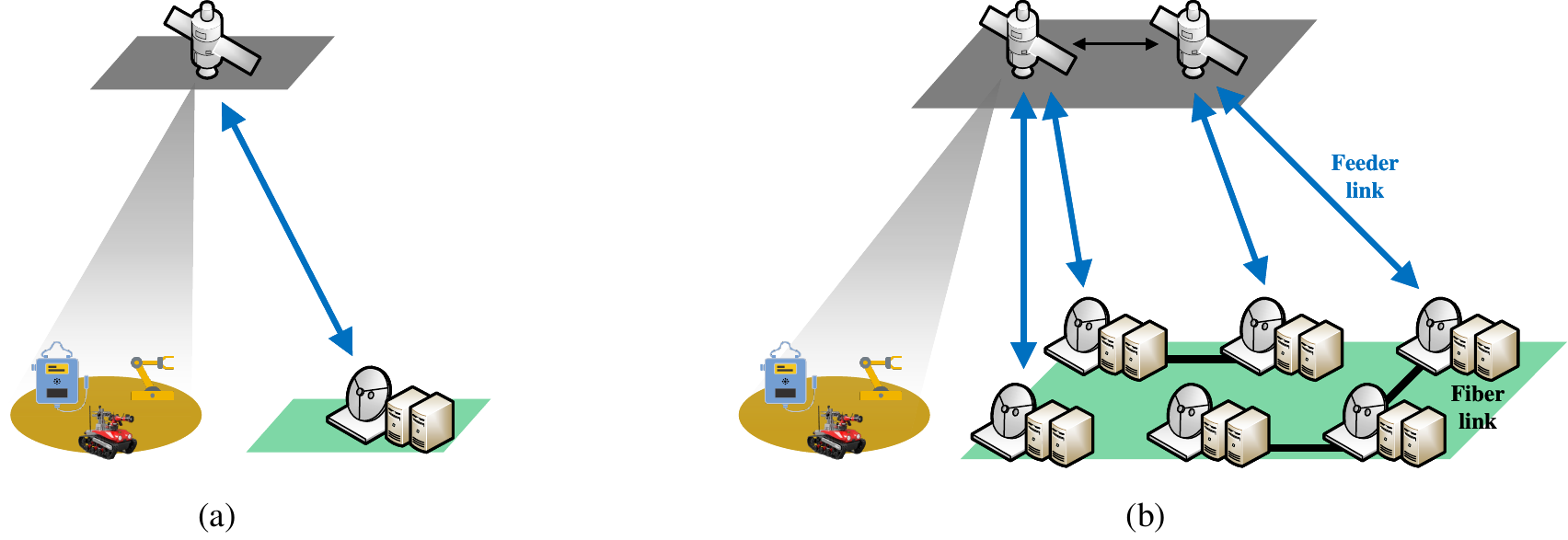}%
		\label{CAF basic}}
	\caption{(a) Illustration of the basic CAF structure \cite{mag00}. (b) Illustration of one possible extension of the CAF structure.}
\end{figure*}


\section{Computing-after-Feeder-link Structure}
The third minimal structure is the CAF structure. As shown in Fig. 3(a), this minimal integrating structure consists of a satellite, a gateway equipped with an MEC server and multiple IoT devices. The minimal structure can be extended by considering multiple gateways, as shown in Fig. 3(b).

In this minimal structure, the MEC servers have higher CPU capability, but the time delay for the IoT devices to access the servers is also higher. Therefore, the CAF structure is suitable for wide-area computation-intensive but delay-tolerant applications. Researches on the CAF structure have mainly focused on computation offloading.

\subsection{Computation Offloading}
The authors of \cite{J-Sup-2} considered a system consisting of multiple LEO satellites and a gateway station equipped with an MEC server. To achieve fast and energy efficient offloading, the bandwidth and power resources of users were jointly allocated. The authors introduced a multi-agent architecture where each LEO satellite made their own allocation policies based on historical policies, as well as users’ workload situation provided by an information center. Based on that, a novel multi-agent information broadcasting and judging algorithm was proposed to allocate resources in a collaborative manner.

\subsection{Lessons Learned and Future Directions}
There are only a few studies that investigate the CAF structure, which mainly focus on the resource allocation in computation offloading.

For the CAF structure, one possible future direction is the problem of deciding the MEC capability configured for the gateway stations. Since gateways often have sufficient energy provision, the main focus of this problem is to satisfy users’ service requirements. This problem can be difficult since that we need to consider not only the service requirement number of the gateways’ neighboring areas, but also farther areas that satellites may cover.

As far as the computation offloading is concerned, existing studies \cite{J-Sup-2} only investigated the resource allocation problem. In fact, the offloading decision is also important in this structure, which decides which gateway station the tasks should be offloaded. The offloading may involve ISL transmissions, which makes this problem even more complicated.

\section{Combination of Minimal Structures}
After discussing the three minimal structures, possible combinations of the minimal structures and the relevant problems will be discussed in this section.

\begin{table*}[b]
	\caption{Summary of Advances on the Combination of CIF and COO Structure}
	\centering
	\renewcommand{\arraystretch}{2}
	\begin{tabular}{|p{0.65in}|p{0.3in}|p{1.2in}|p{1in}|p{3in}|}
		\hline
		\textbf{Theme} & \textbf{Ref.} & \textbf{Network architecture} & \textbf{Design objective}  & \textbf{Proposed solution} \ \\ \hline
		\multirow{6}{0.5in}{Computation offloading} & \cite{C-016} & An LEO satellite and a UAV & Energy & Jointly optimize the task offloading decision and UAV trajectory by proposing an alternating algorithm based on successive convex approximation. \\ \cline{2-5}
		~ & \cite{C-026} & An LEO satellite and multiple UAVs & Latency & Jointly design the offloading decision and resource allocation, and propose a solution to the problem based on problem decomposition and the block coordinate descent method. \\ \cline{2-5}
		~ & \cite{C-065} & An LEO satellite and multiple UAVs & Energy & Jointly optimize the task offloading decision and resource allocation by proposing a low-complexity algorithm based on successive convex optimization. \\ \cline{2-5}
		~ & \cite{J-042} & An LEO satellite and multiple HAPs & Energy & Jointly design the task offloading decision, network resource allocation and MIMO transmit precoding, and propose an algorithm to decompose the problem and solve the sub-problems iteratively.  \\ \cline{2-5}
		~ & \cite{C-213} & LEO constellation with ISLs and a ground base station & Latency/Energy & Optimize the task offloading decision by proposing a double edge computation offloading algorithm based on the Hungarian algorithm. \\ \cline{2-5}
		~ & \cite{J-069} & LEO constellation and a UAV & Latency & Optimize the task offloading decision by proposing a deep reinforcement learning based algorithm. \\ \hline
		Service placement and computation offloading & \cite{C-034} & An LEO satellite and multiple ground base stations & Latency, resource utilization and service caching ratio & Jointly design the service placement strategy, offloading decision and resource allocation by introduced the non-dominated sorting genetic algorithm II. \\ \hline
		Content delivery & \cite{C-214} & An LEO satellite and multiple ground base stations & / & Propose a novel cooperative multicast-unicast transmission scheme to handle both the popular requests and the personalized requests. \\ \hline
	\end{tabular}
	\label{CIF-COO}
\end{table*}

\subsection{Combination of CIF and COO}
In this section, we review the existing studies focused on a combined CIF and COO structure.

A major part of the existing works focused on the computation offloading problem. In \cite{C-016}, the authors considered a system consisting of a UAV and an LEO satellite, both of which were equipped with an MEC server. The joint task offloading decision and UAV trajectory design problem was investigated to minimize the total energy consumption. The authors proposed an alternating algorithm based on the successive convex approximation approach to solve the problem. Other works considered a more complicated system, consisting of a satellite and multiple UAVs or high-altitude platforms (HAPs) with MEC server equipment. The authors of \cite{C-026} considered a latency-oriented joint offloading decision and resource allocation problem in the network. The authors proposed a solution to the problem by decomposing the problem and utilizing the block coordinate descent method. In \cite{C-065}, the authors also investigated the task offloading decision and resource allocation in this network, but turned to minimize the total power consumption by the satellite, UAVs and users. A low-complexity algorithm based on successive convex optimization was proposed to solve the problem. In \cite{J-042}, the system consisted of an LEO satellite and multiple HAPs. Specially, the user-HAP and HAP-LEO communication links all adopted multiple-input-multiple-output (MIMO) techniques. In this paper, the task offloading decision and the network resource allocation were jointly designed with the MIMO transmit precoding. The aim was to minimize the total energy consumption of communication and computation in the system. The authors proposed an algorithm to decompose the problem and solve the sub-problems iteratively. In addition, some studies considered the scenario of utilizing multiple satellites in space. In the system model of \cite{C-213}, users offloaded their task data to a ground base station for edge computing. The data could further be offloaded to an access satellite and transmitted to other satellites through ISLs for data processing. To allocate users’ tasks, the authors proposed a double edge computation offloading algorithm based on the Hungarian algorithm. This proposed algorithm could minimize respectively the average task latency and the average energy consumption of edge servers. In \cite{C-0417-010}, an HAP collected and processed users' task data. Distinguishing from \cite{C-213}, the HAP could further offload the task data to multiple satellites simultaneously for edge computing. The task offloading decision and the communication and computing resource allocation were jointly optimized to achieve energy-minimization in the system. The authors decoupled the problem and proposed an intelligent heuristic algorithm for solution. Moreover, \cite{J-069} jointly considered the CIF and COO combined network with the terrestrial MEC network. Specifically, a UAV with MEC equipment collected users’ computation tasks. The tasks could be executed at the UAV-based server, or offloaded to ground-based or satellite-based servers. The optimization of the offloading decision was performed to minimize the average execution latency. The problem was formulated into a Markov decision process, which was solved by a deep reinforcement learning based algorithm. 

The authors of \cite{C-034} took a step further to jointly consider service placement and computation offloading in a combined CIF and COO network. Specifically, some users in the system offloaded not only their task data but also the corresponding execution codes. The execution codes were cached in ground-based or satellite-based servers, which could then handle offloaded tasks of the same service type. The service placement strategy, offloading decision and resource allocation were jointly optimized in the network. The aim was to minimize the system cost, which was a weighted sum of task latency, computation resource utilization, bandwidth utilization and cache ratio. The authors introduced the non-dominated sorting genetic algorithm II to solve the problem.

The authors of \cite{C-214}, on the other hand, focused on the content delivery problem. They considered a network where MEC servers were placed on the satellite and ground base stations. A novel cooperative multicast-unicast transmission scheme was proposed to handle both the popular requests and the personalized requests. Table \ref{CIF-COO} gives a summary of all these works.

\begin{table*}[b]
	\caption{Summary of Advances on the Combination of COO and CAF Structure}
	\centering
	\renewcommand{\arraystretch}{2}
	\begin{tabular}{|p{0.65in}|p{0.3in}|p{1.2in}|p{1in}|p{3in}|}
		\hline
		\textbf{Theme} & \textbf{Ref.} & \textbf{Network architecture} & \textbf{Design objective}  & \textbf{Proposed solution} \ \\ \hline
		\multirow{9}{0.5in}{Computation offloading} & \cite{C-127} & An LEO satellite and a gateway station & Latency and energy & Jointly optimize the task offloading decision and the bandwidth allocation by proposing a deep reinforcement learning-based algorithm. \\ \cline{2-5}
		~ & \cite{C-045} & An LEO satellite and a gateway station & Latency & Jointly optimize the task offloading decision and allocation, and solve the problem by leveraging the framework of Lyapunov optimization to convert the problem into multiple sub-problems. \\ \cline{2-5}
		~ & \cite{C-103} & LEO constellation and a gateway station & Latency and energy & Jointly optimize the task offloading decision and bandwidth allocation, and solve the problem in two stages by a distributed deep learning algorithm. \\ \cline{2-5}
		~ & \cite{C-036} & LEO constellation and a gateway station & Latency, energy and resource utilization & Jointly optimize the task offloading decision and computation resource allocation by adopting a game-based perspective, and propose a hybrid particle swarm optimization-based algorithm to achieve the Nash equilibrium.  \\ \cline{2-5}
		~ & \cite{C-007} & Three LEO satellites with ISLs and a gateway station & Energy & Jointly optimize the task offloading decision and computation resource allocation based on the improved non-dominated sorting genetic algorithm II. \\ \cline{2-5}
		~ & \cite{C-038} & LEO constellation with ISLs and a gateway station & Latency and energy & jointly optimized the task offloading decision and computation resource utilization by designing a deep reinforcement learning method based on proximal policy optimization. \\ \cline{2-5}
		~ & \cite{C-056} & LEO constellation with ISLs and a gateway station & Energy & Optimize the inter-satellite routing scheme jointly with the task offloading decision and transmission power, and solve the problem by a two-stage algorithm. \\ \cline{2-5}
		~ & \cite{J-028} & LEO constellation with ISLs and a gateway station & Latency and resource utilization & Jointly optimize the task offloading decision and the communication resource utilization, and solve the problem by decomposing it into two sub-problems.  \\ \cline{2-5}
		~ & \cite{J-094} & LEO constellation and multiple gateway stations & Latency and energy & Jointly design the task offloading decision and resource allocation by proposing a A solution based on deep reinforcement learning. \\ \hline
		Computation offloading and content delivery & \cite{J-048} & LEO constellation and GEO/MEO with ISLs, and a gateway station & Latency and resource utilization & Jointly Design the task offloading decision and caching decision by proposing a deep imitation learning-driven offloading and caching algorithm to achieve real-time decision making. \\ \hline
	\end{tabular}
	\label{COO-CAF}
\end{table*}

\subsection{Combination of COO and CAF}
In this section, we review the existing works considering a combined COO and CAF structure.

We first summarize the studies focused on the computation offloading problem. Some studies considered a simple system model which consisted of an LEO satellite and a ground gateway station, both equipped with an MEC server. The authors of \cite{C-127} jointly optimized the task offloading decision and the bandwidth allocation of user-satellite and satellite-gateway links, to minimize the weighted sum of task execution latency and energy consumption. The authors proposed a deep reinforcement learning-based algorithm to solve the problem, which could achieve near-optimal offloading cost performance with low computation complexity. In \cite{C-045}, joint optimization of task offloading decision and resource allocation was performed to minimize the time-averaged task execution latency. The authors leveraged the framework of Lyapunov optimization to convert the problem into multiple sub-problems, which were then solved in an iterative manner.

Further studies were conducted which focused on utilizing multiple satellites in space and a single gateway station. Assuming each user was associated with a single satellite, the authors of \cite{C-103} jointly optimized the offloading decision and bandwidth allocation, aimed at both latency and energy costs. A distributed deep learning algorithm was introduced to solve the problem in two stages. Adopting the same user association scheme, \cite{C-036} investigated the computation offloading of two types of computation tasks, namely edgy-cloud and cloudy-edge. Joint optimization of offloading decision and computation resource allocation were considered for each user to reduce the system costs. Specifically, the system costs took delay, energy consumption, and resource utilization into consideration. The authors provided a game-based perspective on the problem and proposed a hybrid particle swarm optimization-based algorithm to achieve the Nash equilibrium. Some studies further included ISLs in the system model, where task data could be transmitted between satellites through ISLs. In \cite{C-007}, the authors considered a simple scenario with an access satellite and two nearby satellites. Joint task offloading decision and computation resource allocation were conducted for energy consumption minimization. The authors provided a solution to the problem based on the improved non-dominated sorting genetic algorithm II. A more complicated system model which utilized an LEO constellation with ISLs in space was further explored \cite{C-038}. The authors jointly optimized the task offloading decision and computation resource utilization, aimed at lowering both the task execution latency and energy costs. A deep reinforcement learning method based on proximal policy optimization was designed to approximate the optimal solution. Considering a similar network, the authors of \cite{C-056} optimized the inter-satellite routing scheme, jointly with the task offloading decision and the transmission power. The objective was to minimize the energy consumption at the satellites while fulfilling latency constraints. The authors proposed an algorithm which decomposed the problem and solved it in two stages. In \cite{J-028}, a computation task was modeled as a directed graph consisting of multiple virtual network functions. These virtual network functions could be uploaded to different satellites through user-satellite links or ISLs for execution. Joint optimization of offloading decision and communication resource utilization was conducted for bandwidth and delay cost minimization. A distributed algorithm based on multi-agent systems was proposed, which achieved better system performance than the Viterbi and game theory algorithms.

Moreover, a more complicated system could be considered, which consisted of multiple LEO satellites and multiple gateway stations, each with an attached MEC server \cite{J-094}. The authors investigated the joint task offloading decision and resource allocation problem in the network, with the aim of improving system latency and on-orbit computing energy consumption. A solution based on deep reinforcement learning was proposed for this problem.

On the other hand, the authors of \cite{J-048} jointly considered the computation offloading and content delivery problem. Specifically, the computation tasks could be executed on satellites or at the gateway, and the results could be cached on satellite-based servers for further reuse. Joint optimization of task offloading decision and caching decision was performed, aiming to improve both the latency performance and the resource utilization in the system. To this end, the authors proposed a deep imitation learning-driven offloading and caching algorithm which could achieve real-time decision making. Table \ref{COO-CAF} summarizes and compares these works.

\subsection{Combination of CIF and CAF}
Very few studies have focused on the combination of the CIF structure and the CAF structure. The authors of \cite{J-0417-007} considered a system with multiple ground base stations, a satellite and a gateway, where MEC servers were deployed at the base stations and the gateway. The users' task data could be offloaded to an associated base station, or further to the distant gateway with stronger computing capability through satellite communication. The task offloading decision was jointly optimized with the communication and computation resource allocation to minimize the task execution delay. The authors divided the problem into to sub-problems, where the task offloading decision subproblem was solved with theoretical analysis and mathematical derivation, and the resource allocation problem was solved by utilizing the particle swarm optimization algorithm.

\begin{figure*}[tb]
	\centering
	{\includegraphics[width=7in]{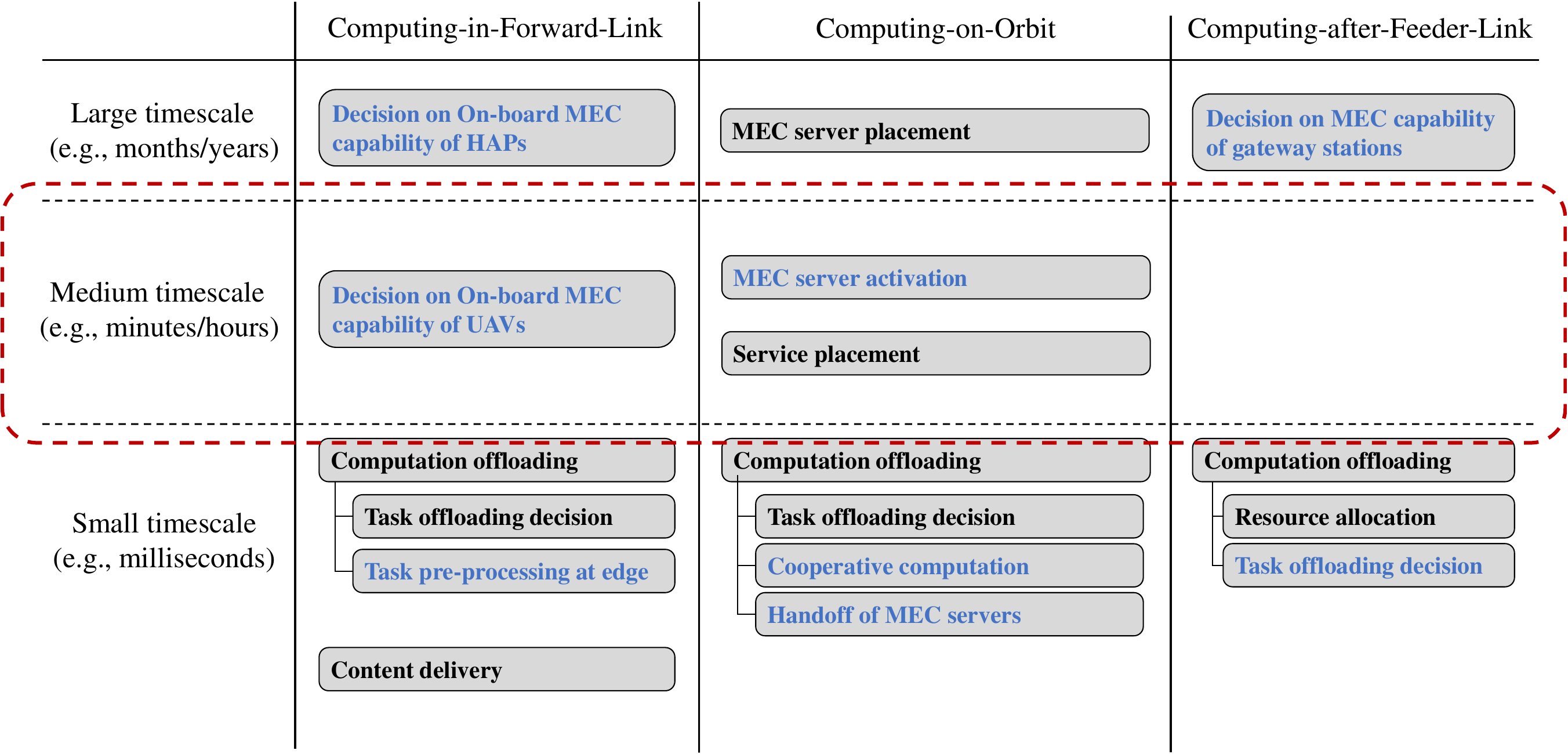}%
		\label{discuss}}
	\caption{Existing and future research directions of the three basic structures at different timescales. (The blue-colored topics are yet to be discussed.)}
\end{figure*}

\section{Open Research Issues}
This section outlines a few open research issues in the integration of satellite and MEC.

\subsection{Hierarchical Orchestration of Minimal Structures in the Integrated Satellite-MEC Network}
In this paper, we provided a new perspective where the integrated satellite-MEC network can be viewed as a nonlinear orchestration of three minimal structures. However, the scale of a practical integrated satellite-MEC network is often huge, and the network structure is complicated. Therefore, it is an important while challenging problem to design an efficient orchestration framework for the minimal structures. Inspired by the structure of proteins, we believe that adopting a hierarchical minimal structure orchestration framework could be a promising solution. Specifically, minimal structures can be simply orchestrated into secondary structures (amino acids orchestrated into peptides), which further form into more complicated tertiary structures (larger peptides), and so forth. Eventually, these structures form into highly functional integrated satellite-MEC networks (proteins). In this framework, the system design of a tertiary structure, for instance, could directly utilize the secondary structures as basic elements for network orchestration, without going into the details of the lower-level minimal structures. Therefore, the computation complexity of network orchestration could be significantly reduced. Further researches into this hierarchical orchestration framework could be considered.

\subsection{Hierarchical-Timescale Network Adjustments in the Integrated Satellite-MEC Network}
The current network is mainly adjusted or optimized at two timescales. Network planning and network architecture adjustment often take place at a large timescale ({\it e.g.}, months or years). Specific communication and computation parameters are adjusted at a small timescale ({\it e.g.}, milliseconds). For the integrated satellite-MEC network, service requirements’ number, service type and spatial distribution change dynamically at a timescale in between. Thus, the traditional network adjustment is unable to match the service requirements, resulting in degraded resource efficiency. Therefore, a hierarchical-timescale network adjustment framework is of interest. We categorize in Fig. 4 the existing and future research topics, based on a hierarchical-timescale framework. It can be observed that for each minimal structure there exist many problems of medium-timescale network adjustment to be explored. Moreover, since the medium timescale is usually much larger than the channel coherence time, a process-oriented optimization framework could be considered for the design of network adjustments \cite{proc ori}.

\subsection{AI-based Integrated Satellite-MEC Network}
AI-based tools and methods have been widely used recently. AI based methods can be applied to the integrated satellite-MEC network in two aspects. First, the integration of satellite and MEC raises new problems, some of which may be hard to model. In this context, AI-based methods can be utilized to provide a feasible solution. On the other hand, the widely distributed MEC servers with close proximity to users can support AI-based applications in return. In fact, \cite{C-102} considered utilizing MEO and LEO satellites to implement a federated learning network. Further investigations into both aspects can be considered.

\subsection{Security Issues in the Integrated Satellite-MEC Network}
Security is an important issue for the integrated satellite-MEC network. Satellite networks provide coverage for a wide geographical area. The openness of electromagnetic environment makes the network susceptible to cyber-attacks of different types, such as eavesdropping and jamming. Besides, the sophisticated integration of satellite and MEC recalls novel system design methods, which may also raise new security risks. To address these problems, new security measures need to be designed and implemented in the network. One possible solution is to combine the integrated satellite-MEC network with the blockchain technique, where each MEC server can work as a node in the blockchain network. However, this may require massive data transmission for data synchronization, which can be difficult for the integrated satellite-MEC network. This tradeoff between security and resource utilization can be further investigated.

\subsection{Integrated Satellite-MEC Network Coordinated with Sensing and Navigation}
Satellites not only play an important role in communication, but also have many other functions such as positioning, navigation and remote sensing. In fact, these functions also produce a substantial amount of data to be transmitted and/or processed. This leads to a new research topic of coordinating the integrated satellite-MEC network with these functions. The work \cite{C-096} has considered a simple coordination scenario where the computation tasks generated by the satellite itself and offloaded by ground users are processed together. This coordination problem can be further investigated to achieve functional cooperation in an efficient manner. With efficient coordination, it is expected that the resource efficiency can be further improved, and new applications that require joint sensing-communication-computation capabilities can be created.

\subsection{Green Integrated Satellite-MEC Network}
In the integrated satellite-MEC network, a huge number of MEC servers will be deployed in a hierarchical manner to provide services, which leads to massive energy consumption. Besides, UAVs, HAPs and other vehicles (automated or manual) are widely adopted in the integrated satellite-MEC network, which also leads to substantial energy consumption and carbon emissions. Therefore, it is important to develop a green integrated satellite-MEC network. Traditional methods used in the terrestrial networks may not be applicable, because servers and vehicles in the integrated network are highly mobile and distributed sparsely and heterogeneously in wide area. Novel techniques for a greener network can be interesting.

\section{Conclusions}
In this paper, we have captured the latest technical advances in satellite-MEC integration. In particular, we have first discussed the main challenges for integrated satellite-MEC networks’ system design, and we have introduced three minimal integrating structures to cope with these challenges. Specifically, the complex integrated network can be regarded as an extension and combination of the minimal structures, providing a new angle for system design. For each minimal structure, we have presented a comprehensive literature review based on the research topics, and have further discussed the gaps and research directions. In addition, we have also reviewed the studies focusing on a combination of minimal structures. Finally, we have outlined the open issues for satellite-MEC integration, such as introducing a hierarchical-timescale network adjustment framework to improve resource efficiency, and combining the integrated network with AI-based techniques, blockchain-based security measures, as well as sensing and navigation functions.

\vfill

\end{document}